\documentclass[twocolumn]{letter_tmp}
\usepackage{amsmath}
\usepackage{amsthm}
\usepackage{amssymb}
\usepackage{bm}
\usepackage{here}
\usepackage{fancybox}
\usepackage{ascmac}
\usepackage{wrapfig}
\usepackage{graphicx}
\group{%Fill in the corresponding group
Scientific Computation and Numerical Analysis
}
\affiliation{Aoyama Gakuin University}{5-10-1 Fuchinobe, Chuo-ku, Sagamihara-shi, Kanagawa 252-5258, Japan}
\affiliation{Tottori University}{4-101 4-101 Koyama-cho Minami, Tottori,  680-8552, Japan}
\authorinfo{Hisashi Kohashi}{1}{kohashi.hisashi@gmail.com}
\authorinfo{Kosuke Sugita}{1}{ksk.sgt@gmail.com}
\authorinfo{Masaaki Sugihara}{1}{}
\authorinfo{Takeo Hoshi}{2}{hoshi@damp.tottori-u.ac.jp}
\title{
Efficient methods for computing integrals in electronic structure calculations
}
\abstract{
Efficient methods are proposed, for computing integrals appeaing in electronic structure calculations.
The methods consist of two parts: the first part is to represent the integrals as contour integrals and the second one is to evaluate the contour integrals by the Clenshaw-Curtis quadrature.
The efficiency of the proposed methods is demonstrated through numerical experiments. 
}
\email{ksk.sgt@gmail.com}
\begin{document}

\maketitle

\section{Introduction}

In this paper, we propose efficient methods for computing the following 
integrals appearing  
in electronic structure calculations\cite{TAKAYAMA-2006,Hoshi2011}:
\begin{equation}
I(\mu,\tau) \equiv-\dfrac{1}{\pi}\lim_{\eta \downarrow +0}\, 
{\rm Im}\!\int_{-\infty}^{\infty}\!W(x; \mu, \tau) G(x+{\rm i} \eta)\, 
dx,
\label{eq-0}
\end{equation}
where $G(z)$ is the Green's function which is defined as
\begin{equation}
G(z)=\bm{b}^* (zI-H)^{-1}\bm{b},
\label{Green's-1}
\end{equation}
%revision:start------------------------------------------------
%where $\bm{b}$ is a vector and $H$ is the so-called Hamiltonian matrix, 
where $\bm{b}$ is a vector and $H$ is the so-called Hamiltonian matrix (a Hermitian matrix), 
%revision:end--------------------------------------------------
and $W(x; \mu, \tau)$ is the Fermi-Dirac function:
\vspace{-2mm}
\[
W(x; \mu, \tau)=\frac{1}{1+\exp\Big(\dfrac{x-\mu}{\tau}\Big)},
\]
where $\mu$ is a real number, and $\tau$ is a small positive number.
The Green's function $G(z)$ is expanded as follows:
\begin{equation}
G(z)=\sum_{j=1}^N \frac{c_j}{z-\lambda_j},
\label{Green's-2}
\end{equation}
where $N$ is the order of $H$, 
$c_j \: (j=1, 2, \cdots, N)$ are non-negative real numbers, 
and $\lambda_j \:(j=1, 2, \cdots, N)$ are  the eigenvalues (real numbers) 
of $H$, which correspond to the energy levels in the material. 
We assume that $\lambda_j$'s are labeled in increasing order: $\lambda_1<\lambda_2< \cdots<\lambda_N$.
We also assume that a lower estimate for the smallest eigenvalue, $\lambda_1$,
 is known, although $\lambda_j\:(j=1, 2, \cdots, N)$ are unknown. 

In \cite{Hoshi2011}, the trapezoidal rule is applied to evaluate the integral 
in $I(\mu,\tau)$, 
by taking $\eta$ as a very small number, and by setting the interval of 
integration adequately wide.
It is evident that many sampling points are necessary, since $G(z)$ has 
many poles on the real axis.
However, due to a limited time of computation, the number of the sampling 
points is not as many as supposed to be.
Therefore, the accuracy of the computed results is not enough.

In this paper, we propose two methods for computing $I(\mu, \tau)$ 
efficiently, 
both of which consist of two parts: the first part is to represent the integrals as contour integrals and the second one is to evaluate the contour integrals by the Clenshaw-Curtis quadrature.

The paper is organized as follows.
In Sec. 2, we introduce one of the proposed methods, which we call Method 1,  and give a numerical example.
Subsequently, in Sec. 3, we present another method, which we call Method 2, 
together with a numerical example. 
In Sec. 4, we develop a method for computing $I(\mu,\tau)$ 
for many distinct values of $\mu$, which situation sometimes arises.
Finally, in Sec. 5, we make concluding remarks.

\section{
Method 1
}\label{sec::Topdown}

\begin{figure}[H]
  \begin{center}
    \includegraphics[clip,width=6.0cm]{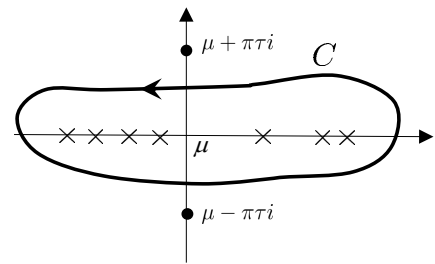}
%\vspace{-1zh}
\vspace{-0.5cm}
    \caption{ The contour $C$ ($\times$: the poles of $G(z)$，$\bullet$: the poles of $W(z; \mu, \tau)$）}
    \label{fig-1}
  \end{center}
\end{figure}

It is easily verified that $I(\mu,\tau)$ is equal to $\displaystyle\sum_{j}c_jW(\lambda_j; \mu, \tau)$.
Thus, by setting the contour $C$ as a simply closed curve which encloses all the poles of $G(z)$ but does not of $W(z)$ (Fig. \ref{fig-1}), the contour integral representation of $I$ is obtained as follows:
\begin{equation}
I(\mu,\tau)=\frac{1}{2\pi i}\int_C W(z; \mu, \tau)G(z) \, dz\,.
\label{eq-1}
\end{equation}
%\vspace{-3.5zh}
\vspace{-1cm}
%\vspace{-2.5zh}
\vspace{-1cm}

It is expected that we will perform the numerical integration efficiently with this contour integral, because we can set the contour far from the poles of $G(z)$ on the real axis.

\begin{figure}[H]
  \begin{center}
    \includegraphics[clip,width=7.0cm]{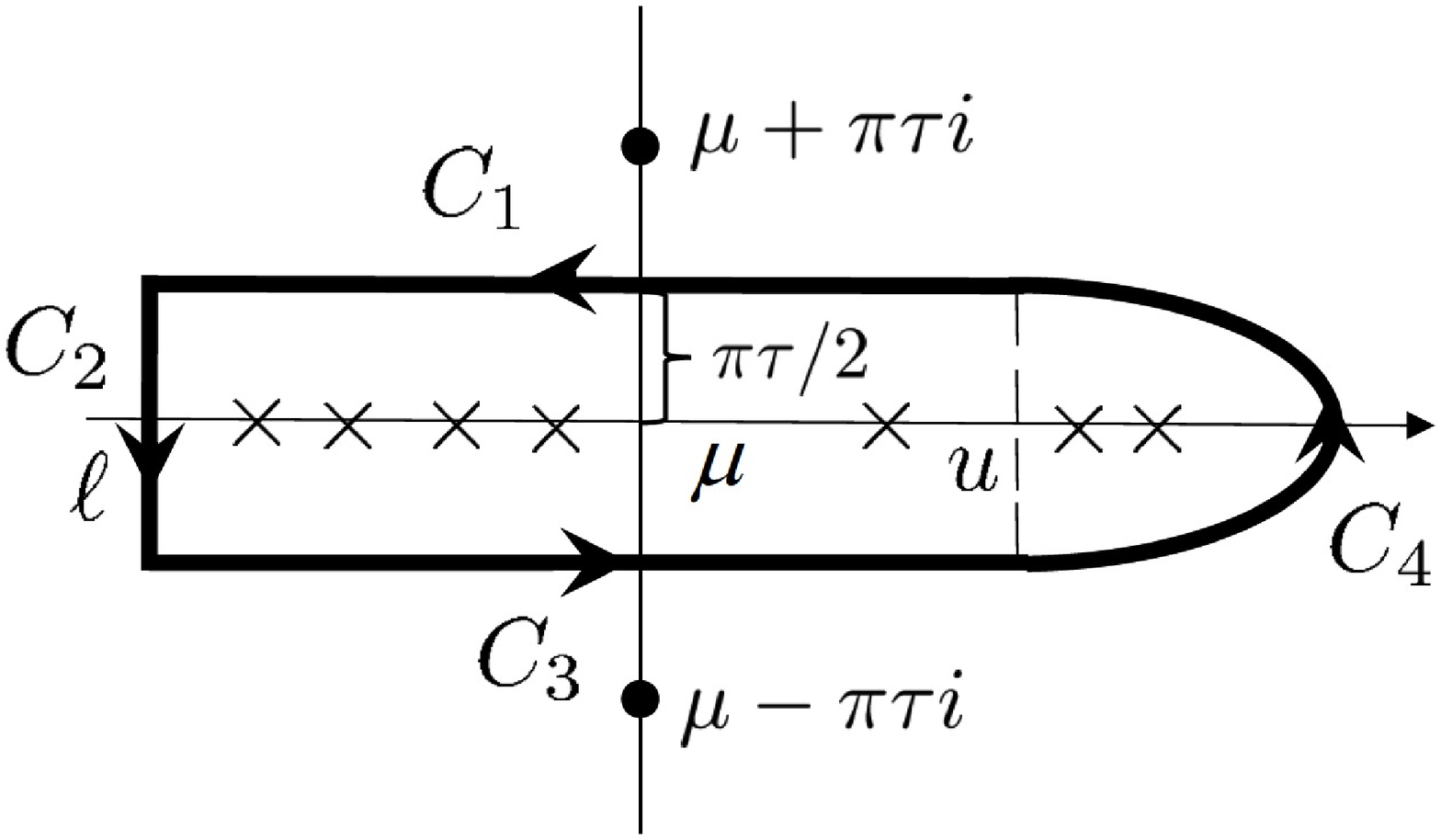}
%\vspace{-2.5zh}
%\vspace{-1cm}
    \caption{The contour $L_1+L_2+L_3+C_4$ 
             ($\times$: the poles of $G(z)$, $\bullet$: the poles of $W(z)$)}
    \label{fig-2}
  \end{center}
\end{figure}

Taking account of easiness of numerical integration, we now set the contour of  the integral as $L_1+L_2+L_3+C_4$ illustrated in Fig. \ref{fig-2}, where
\renewcommand{\labelenumi}{(\alph{enumi})}
%revision:start------------------------------------------------
\begin{enumerate}
\item 
%$\ell$ is a real number that is smaller than the smallest pole of $G(z)$, i.e.,  $\lambda_1$ and such that $W(x; \mu, \tau)\approx 1$ (note that the possibility of setting up this number is guaranteed by the assumption that a lower estimate for $\lambda_1$ is known); 
$\ell$ is a real number that is smaller than the smallest pole of $G(z)$, i.e.,  $\lambda_1$ 
(note that the possibility of setting up this number is guaranteed by the assumption that a lower estimate for $\lambda_1$ is known), 
and also such that $W(\ell; \mu, \tau) \approx 1$, i.e.,  
$|W(\ell; \mu, \tau) - 1|$ is small enough (in the numerical examples below, 
we set $|W(\ell; \mu, \tau) - 1| \le 10^{-40}$);

\item $u$ is a real number such that $W(u; \mu, \tau) \approx 0$ (in the numerical examples below, 
we set $|W(u; \mu, \tau)| \le 10^{-40}$);

\item $L_1=[u+(\pi \tau/2) i, \: \ell+(\pi \tau/2) i]$; 

\item $L_2=[\ell+(\pi \tau/2) i, \: \ell-(\pi \tau/2) i]$; 

\item $L_3=[\ell-(\pi \tau/2) i, \: u-(\pi \tau/2) i]$; 

\item $C_4$ is a curve connecting the points $u-(\pi \tau/2) i$ and $u+(\pi \tau/2) i$ and such that $L_1+L_2+L_3+C_4$ encloses all the poles of $G(z)$ .
\end{enumerate}
%revision:end--------------------------------------------------
%\vspace{-3.5zh}
\vspace{-1cm}
%\vspace{-1.5zh}
\vspace{-0.5cm}

Then, denoting $W(z; \mu, \tau)G(z)$ by $F(z; \mu, \tau)$, we have 
\begin{align*}
&\dfrac{1}{2\pi i}\int_{L_{1}}  F(z; \mu, \tau) dz+
\dfrac{1}{2\pi i}\int_{L_{3}}  F(z; \mu, \tau) dz \\
&\hspace{2.0cm}=-\dfrac{1}{\pi}\:\int_\ell^u {\rm Im}\, F(x+\frac{\pi\tau}{2}i; \mu, \tau)
dx,
\end{align*}
\begin{align*}
\frac{1}{2\pi i}\int_{L_{2}} F(z; \mu, \tau)dz
%%revision:start------------------------------------------------
%&=-\frac{1}{ \pi }\int_{0}^{\frac{\pi\tau}{2}} {\rm Re}\,G(\ell+yi)dy, \\
&\approx -\frac{1}{ \pi }\int_{0}^{\frac{\pi\tau}{2}} {\rm Re}\,G(\ell+yi)dy, \\
%%revision:end--------------------------------------------------
\frac{1}{2\pi i}\int_{C_{4}}  F(z; \mu, \tau) dz 
&\approx
0\,.
\end{align*}
Thus,  we obtain 
\[
I(\mu, \tau)\approx I_{\rm h}+I_{\rm v}
\]
where 
\begin{align}
I_{\rm h}&=-\dfrac{1}{\pi}\int_l^u{\rm Im}\,F(x+\frac{\pi\tau}{2}i; \mu, \tau)dx, \label{eq-2}\\
I_{\rm v}&=-\dfrac{1}{ \pi }\int_{0}^{\frac{\pi\tau}{2}}{\rm Re}\, G(\ell+yi)dy\,. \label{eq-3}
\end{align}
For calculation of $I_{\rm h}$ and $I_{\rm v}$ we adopt the Clenshaw-Curtis quadrature \cite{Trefethen2013},  
because the integrands are analytic over the intervals of integration.

%\vspace{1zh}
\vspace{1cm}

\noindent
{\bf Numerical example 1} \  We consider the case where the Green's function is given by  
\[
G(z)=\sum_{j=1}^{4686} \frac{1}{z-\lambda_j},
\]
which appears in an electronic structure calculation of a nanoscale amourphous-like conjugated polymer\cite{Hoshi2012}．
The values of $\lambda_1, \lambda_2, \cdots , \lambda_{4686}$
($\lambda_{1}\simeq-1.16, \lambda_{4686}\simeq5.58$)
are given on the website \cite{Elses} as the data set ^^ ^^ APF4686''. 
(The actual computation of $G(z)$ is done by using the expression (\ref{Green's-1}), that is, by solving the large system of linear equations $(zI-H)\bm{x}=\bm{b}$, which leads to relatively large numerical errors.
Since we here concentrate on examining numerical errors caused by the numerical integration, we give $G(z)$ as the rational expression as above, 
the computation of which produces small numerical errors.)

\begin{figure}[H]
\begin{center}
      \includegraphics[width=7cm,clip]{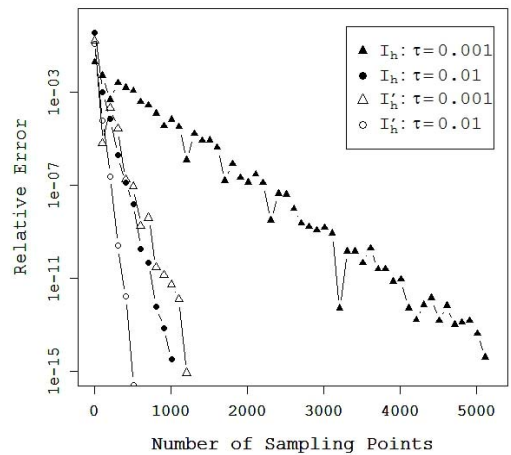}
%\vspace{-1zh}
\vspace{-0.5cm}
         \caption{The convergence behaviors of the Clenshaw-Curtis quadrature 
for $I_{\rm h}$ and $I_{\rm h}'$}
\label{fig-3}
\end{center}
\end{figure}

We first set $\mu$ as 
\begin{equation}
\mu=(\lambda_{2343}+\lambda_{2344})/2=-0.3917431575916144,
\label{mu-1}
\end{equation}
where 
\begin{align*}
  \lambda_{2343}=&-0.4258775547956950,\\ 
  \lambda_{2344}=&-0.3576087603875338,
\end{align*}
and $\tau=0.01$.
In this case, the exact value of $I(\mu, \tau)$ is $2342.992785654893\cdots$.
We evaluate the integrals $I_{\rm h}$ and $I_{\rm v}$
%%revision:start------------------------------------------------
with the Clenshaw-Curtis quadrature, setting $\ell=-1.5,u=0.6$.
%with the Clenshaw-Curtis quadrature, setting $\ell=-1.5,u=0.6$ 
%so that $|W(\ell;\mu,\tau) - 1| \approx \exp(-110) \approx 10^{-48}$ and $|W(u;\mu,\tau)| \approx \exp(-100) \approx 10^{-44}$.
%%revision:end--------------------------------------------------
For $I_{\rm v}$, whose integrand  has no poles near the interval of integration, a very rapid convergence of the Clenshaw-Curtis quadrature is observed: 
the relative error $10^{-15}$ is attained with $7$ sampling points.
For $I_{\rm h}$, whose integrand  has many poles near 
the interval of integration,  the convergence behavior is shown 
in Fig. \ref{fig-3}.
Exponential convergence is observed, 
as expected from the convergence theory of 
the Clenshaw-Curtis quadrature.
%\vspace{-3zh}
\vspace{-1.0cm}
%\vspace{-2.5zh}
Next, we set $\mu$ the same as above and $\tau=0.001$.
The exact value of $I(\mu, \tau)$ is $2343.000000000000\cdots$.
We evaluate $I_{\rm h}$ and $I_{\rm v}$ with the Clenshaw-Curtis quadrature, 
%%revision:start------------------------------------------------
setting $\ell=-1.5,u=-0.3$.
%setting $\ell=-1.5,u=-0.3$ so that $|W(\ell;\mu,\tau) - 1| \approx \exp(-1100) \ll 10^{-48}$ and $|W(u;\mu,\tau)| \approx \exp(-100) \approx 10^{-44}$.
%%revision:end--------------------------------------------------

For $I_{\rm v}$, the relative error $10^{-15}$ is achieved 
with $6$ sampling points.
For $I_{\rm h}$, Fig. \ref{fig-3} shows the convergence behavior. 
Exponential convergence is observed, which is slower than that of the case of $\tau=0.01$.

\section{
Method 2
}

\begin{figure}[H]
  \begin{center}
    \includegraphics[clip,width=7.0cm]{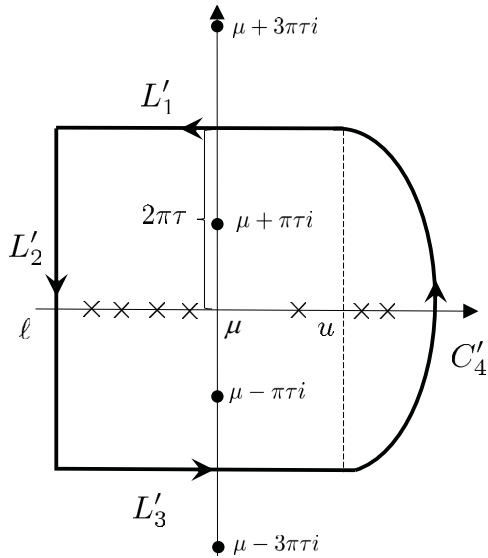}
    \caption{The contour $L'_1+L'_2+L'_3+C'_4$ ($\times$: the poles of $G(z)$, 
$\bullet$: the poles of $W(z)$）}
    \label{fig-4}
  \end{center}
\end{figure}

Fig. \ref{fig-3} shows that Method 1 requires a large number of sampling points, that is, function evaluations for computing the integral $I_h$, which caused by the proximity of the paths $L_1, L_3$ to the poles of $W(z)G(z)$.
To solve this problem, we take the contour $L'_1+L'_2+L'_3+C'_4$ so that it is far from the poles of $W(z)G(z)$, as shown in Fig.\ref{fig-4}.

%\vspace{-1.5zh}
%\vspace{-0.75cm}
%\vspace{-1.5zh}
\vspace{-0.75cm}
In this setting, we should consider the residues of $W(z; \mu, \tau)$ at $z=\mu\pm \pi \tau i$, thus the contour integral becomes
\begin{align*}
&\dfrac{1}{2 \pi i }\int_{L'_1+L'_2+L'_3+C'_4} F(z; \mu, \tau)dz \\
& \hspace{1cm} =I(\mu, \tau)+{\rm{Res}}(F, \mu+\pi \tau i)+{\rm{Res}}(F, \mu-\pi \tau i)\\
& \hspace{1cm} =I(\mu, \tau)-2 \tau \: {\rm Re} \, G(\mu+\pi \tau i).
\end{align*}
Calculating the left-hand side similarly as in Method 1, we obtain
\[
I(\mu, \tau) \approx I_{\rm h}'+I_{\rm v}'+ 2 \tau \: {\rm Re} \, 
G(\mu+\pi \tau i),
\] 
where 
\begin{align}
I_{\rm h}'&=-\dfrac{1}{\pi}\int_l^u {\rm Im}\, 
F(x+2\pi\tau i; \mu, \tau)dx \label{eq-4},\\
I_{\rm v}'&=-\dfrac{1}{ \pi }\int_{0}^{2\pi\tau} {\rm Re}\, G(l+yi)dy. 
\label{eq-5}
\end{align}
For evaluation of $I_{\rm h}'$ and $I_{\rm v}'$,
we use the Clenshaw-Curtis quadrature.

%\vspace{1zh}
\vspace{0.5cm}

\noindent
{\bf Numerical example 2} \ 
We set $G(z), \mu$, and $\tau$ as the same as in Numerical example 1.
First, in the case of $\tau=0.01$, we evaluate 
the integrals $I_{\rm h}'$ and $I_{\rm v}'$ with the Clenshaw-Curtis quadrature, setting $\ell=-1.5,u=0.6$. 
For $I_{\rm v}'$, the relative error $10^{-15}$ is attained with $12$ 
sampling points.
For $I_{\rm h}'$, Fig. \ref{fig-3} shows the convergence behavior of 
the Clenshaw-Curtis quadrature.
Next, in the case of $\tau=0.001$, we compute the integrals $I_{\rm h}'$ and 
$I_{\rm v}'$ with the Clenshaw-Curtis quadrature, setting $\ell=-1.5,u=-0.3$.
For $I_{\rm v}'$ the relative error $10^{-15}$ is achieved 
with $6$ sampling points.
For $I_{\rm h}'$, Fig. \ref{fig-3} shows the convergence behavior of 
the Clenshaw-Curtis quadrature.
We can see that the performance is much improved.

%\vspace{1zh}
\vspace{0.5cm}

\noindent 
{\bf Remark 1} \ Setting such contour contributes not only to fast computations of integrations, but also to the actual computation of $G(z)$.
In fact, the actual computation of $G(z)$ requires to solve 
the equation $(zI-H)\bm{x}=\bm{b}$ with Krylov subspace methods such as the COCG method. And it is known that the farther 
the distances between $z$ and the eigenvalues of $H$, i.e., $\lambda_j$'s, 
the faster the convergence of Krylov subspace methods in general.

\section{
Method for computing $I(\mu,\tau)$ with various values of $\mu$
}\label{sec::Example}

It is often the case that we need to compute $I(\mu,\tau)$ 
for many distinct values of $\mu$.
Computing separately for each value of $\mu$ with Method 1 or 2, costs a massive amount of calculation in total.
Instead we  propose an efficient method  based on Method 1. 
It is supposed here that the range of required $\mu$, say $[\mu_{\rm min}, \mu_{\rm max}]$, is known.

In Method 1, it is evident that the computation of $I_{\rm h}(\mu, \tau)$ for many values of $\mu$ causes the massive amount of computation. 
Hence we reduce the amount of the computation of $I_{\rm h}(\mu, \tau)$ for many values of $\mu$. 
The key is to use common $\ell$ and $u$ (See (\ref{eq-2})) in the computation of $I_{\rm h}(\mu,\tau)$ for many values of $\mu$. 
In fact, we can use $\ell_{\rm min}$ and $u_{\rm max}$ as the common $\ell$ 
and $u$ respectively, where $\ell_{\rm min}$ is a real number 
that is less than or equal to the value of $\ell$ determined in the case of $\mu=\mu_{\rm min}$,  
and $u_{\rm max}$ is a real number that is greater than or equal to the value 
of $u$ determined in the case of $\mu=\mu_{\rm max}$. Then, 
%\vspace{0.5zh}
\vspace{0.25cm}
\begin{align*}
I_{\rm h}&=-\dfrac{1}{\pi}\int_{\ell_{\rm min}}^{u_{\rm max}}{\rm Im}\:W(x+\frac{\pi\tau}{2}i; \mu, \tau)\,G(x+\frac{\pi\tau}{2}i)dx\,. \\
\Bigg(I_{\rm v}&=-\dfrac{1}{ \pi }\int_{0}^{\frac{\pi\tau}{2}} 
{\rm Re}\, G(\ell_{\rm min}+yi)dy\Bigg) 
\end{align*}
%\vspace{0.5zh}
\vspace{0.25cm}
It follows that the computation of $G$, which costs a large amount of 
calculation, is independent of $\mu$. 
Thus, once we compute $I_h$ for an appropriately large $\mu$ and store the computed values of $G(x+(\pi\tau/2)i)$ for reuse, we can immediately obtain the result of $I_{\rm h}$ for another value of $\mu$, by multiplying the stored values of $G(x+(\pi\tau/2)i)$ by $W(x+(\pi\tau/2)i; \mu, \tau)$, the cost of computation of which is low.
%Taking $\mu_{\rm max}$ seems to be suitable for computing $G$ just once, 
%since it is expected that the convergence is slow.
This device enables us to compute $I(\mu,\tau)$ for many of distinct $\mu$ efficiently.

% ---------------------------------------

\bigskip

\begin{figure}[H]
    \begin{center}
      \includegraphics[width=7.5cm,clip]{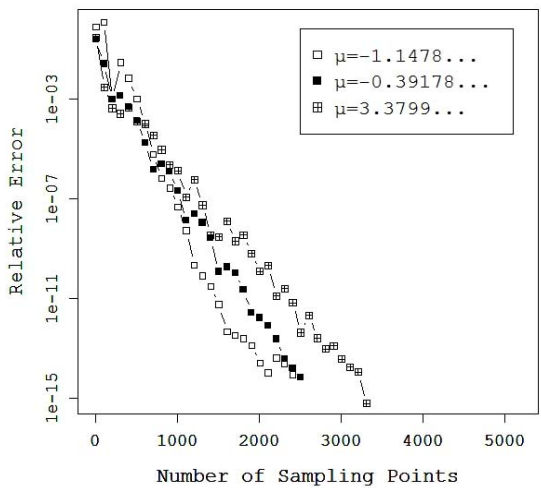}
     \end{center}
\caption{
The convergence behaviors of the Clenshaw-Curtis quadrature for $I_{\rm h}$ (Note that we compute $I_{\rm h}$ for $\mu=(\lambda_{10}+\lambda_{11})/2=-1.1478\cdots$, $\mu=(\lambda_{2343}+\lambda_{2344})/2=-0.39174\cdots$, reusing the value of $G(z)$ obtained by the computation in the case of $\mu=(\lambda_{4650}+\lambda_{4651})/2=3.3799\cdots$)
}
    \label{fig-5}
\end{figure}

\noindent
{\bf Numerical example 3} \ We consider the case where $G(z)$ is the same as
in Numerical example 1 and $\tau = 0.01$.
We compute $I(\mu,\tau)$ for 
\begin{align*}
\mu&=(\lambda_{10}+\lambda_{11})/2=-1.147817562299727,\\
& \hspace{1cm} \vdots\\
\mu&=(\lambda_{4650}+\lambda_{4651})/2=3.379941668485607.
\end{align*}
We apply the Clenshaw-Curtis quadrature to evaluate $I_{\rm h}$ and $I_{\rm v}$ with $\ell_{\rm min}=-1.5,u_{\rm max}=6.0$. 
Note that $I_{\rm v}$ is the same as in Numerical example 1, for which 
the Clenshaw-Curtis quadrature attains the relative error $10^{-15}$ with $6$ sampling points. 
For $\mu=(\lambda_{4650}+\lambda_{4651})/2=3.3799\cdots$, 
we compute $I_{\rm h}$ and store the computed values of $G(z)$.  Then we compute $I_{\rm h}$ for another value of $\mu$, by using the stored value of $G(z)$. 
Fig. \ref{fig-5}  shows the convergence behaviors of 
the Clenshaw-Curtis quadrature for $I_{\rm h}$ with 
$\mu=(\lambda_{10}+\lambda_{11})/2=-1.1478\cdots$,
$\mu=(\lambda_{2343}+\lambda_{2344})/2=-0.39174\cdots$ and 
$\mu=(\lambda_{4650}+\lambda_{4651})/2=3.3799\cdots$.
%\vspace{-3zh}
%\vspace{-1.5cm}
%\vspace{-2.5zh}
\vspace{-1.25cm}

\bigskip

\noindent
{\bf Remark 2} \ As for Method 2, we can use a similar device for computing $I_{\rm h}'$ with various values of $\mu$. 
In fact, once we compute $I_{\rm h}'$ for a suitable $\mu$ and store the values of $G(x+2\pi\tau i)$, then we can obtain the results of $I_{\rm h}'$ for distinct values of $\mu$, by multiplying the stored values of $G(x+2\pi\tau i)$ by 
$W(x+2\pi\tau i; \mu, \tau)$, which does not need much computation.

However, it should be noted that an additional 
computation of $G(\mu+\pi \tau i)$ is needed for the calculation of $I$.
The fact that Method 2 is faster than Method 1, tells us that 
Method 2 with this device can be effective when the total cost of computation 
of $G(\mu+\pi \tau i)$ is not large.

\section{Concluding remarks}\label{sec::Conclusion}

In this paper, we proposed methods for computing integrals appearing 
in electronic structure calculations and showed that the proposed methods are efficient through numerical experiments.
We would like to note that the proposed methods are also efficient for the case where the Green's function is given by 
\begin{align*}
G(z)&=\sum_{j=1}^{4686} \frac{c_j}{z-\lambda_j} \quad 
% \mbox{（$c_j$は$[0,1]$上の一様乱数）}
% \mbox{（$\{c_j\}$: uniform random numbers on $[0,1]$）}
\end{align*}
where $\{c_j\}$ are uniform random numbers on $[0,1]$, although the results are not contained here, due to a limited number of pages.

We applied the proposed methods to the Green's function $G(z)$ 
given as rational expression, but we should treat the Green's function $G(z)$ with (\ref{Green's-1}) using Hamiltonian matrix, which is left for the future work. 

\vspace{5mm}

%\noindent \textbf{Acknowledgement:}
%\textit{ 
%This research was supported by JSPS KAKENHI(Grant-in-Aid for Scientific Researc%h(C)) No. ??????.
%}

%\noindent \textbf{Note:}
%\textit{ The views and opinions expressed here are
%those of the authors and do not reflect the views of the authors employer.}

\references


\begin{thebibliography}{99}
\bibitem{TAKAYAMA-2006}
R. Takayama, T. Hoshi, T. Sogabe, S.-L. Zhang, and T. Fujiwara, 
Linear algebraic calculation of the Green's function for large-scale electronic structure theory,
Phys. Rev. B, {\bf 73} (2006), 165108, 1-9.
\bibitem{Hoshi2011}
H. Teng, T. Fujiwara, T. Hoshi, T. Sogabe, S.-L. Zhang, S. Yamamoto, Efficient and accurate linear algebraic methods for large-scale electronic structure calculations with nonorthogonal atomic orbitals, Physical Review B, {\bf 83} (2011), 165103, 1-12.
\bibitem{Trefethen2013}
Lloyd N. Trefethen, Approximation Theory and Apporoximation Practice, SIAM, Oxford, 2013.
\bibitem{Hoshi2012}
T. Hoshi, S. Yamamoto, T. Fujiwara, T. Sogabe, S.-L. Zhang,
An order-$N$ electronic structure theory with generalized eigenvalue equations 
and its application to a ten-million-atom system, 
J. Phys.: Condens. Matter {\bf 24} (2012), 165502, 1-5.
\bibitem{Elses}
http://www.elses.jp/matrix/ 

\end{thebibliography}
\end{document}